\documentclass[
    journal=jacsat,
    manuscript=article
]{achemso}

\SectionsOn
\SectionNumbersOn
\usepackage[english]{babel}

\graphicspath{{figures/}}

\makeatletter
\def\input@path{{figures/}}
\makeatother

\usepackage{amsmath}
\usepackage{amssymb}
\usepackage{amsfonts}
\usepackage{amsthm}
\usepackage{mathtools}
\usepackage{braket}
\usepackage[version=4]{mhchem}

\DeclarePairedDelimiter\norm{\lVert}{\rVert}
\DeclarePairedDelimiter\abs{\lvert}{\rvert}
\DeclarePairedDelimiter\expval{\langle}{\rangle}

\DeclareUnicodeCharacter{00F6}{\"o}

\usepackage{graphicx}
\usepackage{tikz}
\usepackage{tikz-network}

\usetikzlibrary{
    decorations.pathreplacing,
    decorations.text
}

\usepackage{booktabs}

\usepackage{algorithm}
\usepackage{algpseudocode}

\algrenewcommand\algorithmicdo{}

\usepackage{xcolor}
\usepackage{soul}
\usepackage{etoolbox}
\usepackage[normalem]{ulem}
\usepackage[most]{tcolorbox}

\setkeys{acs}{doi = true}
\newcommand{\doi}[1]{\href{http://dx.doi.org/#1}{\nolinkurl{#1}}}

\usepackage[
    colorlinks=true,
    linkcolor=myblue,
    citecolor=myblue,
    urlcolor=myblue
]{hyperref}

\usepackage{cleveref}

\crefname{appendix}{appendix}{appendices}

\sethlcolor{yellow}

\robustify\ref
\robustify\cite

\soulregister\ref7
\soulregister\cite7
\soulregister\eqref7
\soulregister\ce7

\newcommand{\changed}[1]{#1}

\newtcolorbox{changedblock}{
    breakable,
    colback=yellow!30,
    frame hidden,
    sharp corners,
    boxsep=2pt,
    left=2pt,
    right=2pt,
    top=2pt,
    bottom=2pt
}

\definecolor{myblue}{rgb}{0,0,1}
\definecolor{myred}{rgb}{1,0,0}
\definecolor{mygreen}{rgb}{0,0.6,0}

\newcommand{\redsout}{%
    \bgroup
    \markoverwith{%
        \textcolor{myred}{%
            \rule[0.5ex]{2pt}{0.4pt}%
        }%
    }%
    \ULon
}

\title[
    Efficient AFQMC using ITHC
]{
    Efficient Auxiliary-Field Quantum Monte Carlo
    using Isometric Tensor Hypercontraction
}

\author{Maxine Luo}
\affiliation{
    Max Planck Institute of Quantum Optics,
    Hans-Kopfermann-Stra{\ss}e 1,
    85748 Garching, Germany
}
\alsoaffiliation{
    Munich Center for Quantum Science and Technology,
    Schellingstra{\ss}e 4,
    80799 M\"unchen, Germany
}
\altaffiliation{
    These authors contributed equally to this work.
}

\author{Victor Chen}
\affiliation{
    Technical University of Munich,
    Department of Chemistry,
    Lichtenbergstra{\ss}e 4,
    85748 Garching, Germany
}
\altaffiliation{
    These authors contributed equally to this work.
}

\author{Yu Wang}
\email{18yu.wang@tum.de}
\affiliation{
    Technical University of Munich,
    CIT, Department of Computer Science,
    Boltzmannstra{\ss}e 3,
    85748 Garching, Germany
}

\author{Christian B.~Mendl}
\affiliation{
    Technical University of Munich,
    CIT, Department of Computer Science,
    Boltzmannstra{\ss}e 3,
    85748 Garching, Germany
}
\alsoaffiliation{
    Technical University of Munich,
    Institute for Advanced Study,
    Lichtenbergstra{\ss}e 2a,
    85748 Garching, Germany
}

\begin{document}

\begin{abstract}
Auxiliary Field Quantum Monte Carlo (AFQMC) has emerged as a powerful framework for treating strongly correlated electronic systems, offering a favorable balance between computational cost and accuracy.
In this paper, we present a novel AFQMC method that uses the isometric tensor hypercontraction (ITHC) technique to diagonalize the two-body Coulomb interaction of molecular electronic Hamiltonians by introducing additional fictitious fermionic modes. \changed{While traditional low rank approximations such as the Tensor Hypercontraction (THC) decomposition have been used in AFQMC in the past, their conventional role has been to provide compact representations of the two-electron interaction, thereby reducing the memory footprint. In comparison, our method not only reduces memory usage but also} exhibits lower theoretical complexity and better practical performance for both propagation and local-energy evaluation compared to the standard AFQMC method.
We demonstrate the efficacy of this approach by computing the ground-state energies of a linear $\ce{H10}$-chain and the benzene molecule. 
Our results show that the extended-basis AFQMC recovers many-body correlations with a precision comparable to that of high-level wavefunction methods such as Coupled Clusters (CC) or Density Matrix Renormalization Group (DMRG), while offering significantly improved scaling. 
\end{abstract}

\section{Introduction}
Solving the electronic Schr\"odinger equation lies at the heart of modern computational chemistry and computational material physics. Electron density-based methods such as Density Functional Theory (DFT)~\cite{Kohn.1965}  have seen significant success due to their favorable scaling of $\mathcal{O}(N^3)$ -- $\mathcal{O}(N^4)$ with $N$ being the number of orbitals, which stems from their dependence on the electron density $\rho$ rather than the many-electron wavefunction~\cite{DFT_review}. However, DFT often fails or yields unsatisfactory results when applied to systems with strong electron-electron correlation~\cite{DFT_review, MottaHchain}. While correlated methods, such as Configuration Interaction (CI)~\cite{FCI} or Coupled Clusters (CC)~\cite{CC}, can be applied to obtain more accurate results for these correlated systems, these methods are afflicted by their high computational costs with scalings between $\mathcal{O}(N^4)$ -- $\mathcal{O}(N^{10})$ or even exponential depending on the number of included excited states~\cite{Chemistry_review, CC, FCI}.

Auxiliary Field Quantum Monte Carlo (AFQMC) methods offer a favorable balance between computational cost and accuracy and have become increasingly relevant in describing many-electron systems~\cite{Kim.2018, Motta.2018, review_afqmc}. Unlike DFT and traditional wavefunction-based methods, AFQMC is a statistical method that determines the system's electronic ground state via stochastic sampling~\cite{review_afqmc}. Its sole reliance on the many-body Hamiltonian and trial wavefunction, while maintaining a similar scaling as DFT, renders it a potential alternative to traditional quantum-chemical methods~\cite{Lee.2020, Motta.2018, review_afqmc}. AFQMC methods have been shown to describe highly correlated systems such as the Fermi-Hubbard model~\cite{Huang.2017, Xu.2024} and are free of self-interaction errors~\cite{review_afqmc}. While the accuracy of AFQMC methods approaches that of correlated wavefunction-based methods, their computational cost scaling is significantly lower~\cite{review_afqmc}. 

\changed{A major challenge in AFQMC simulations is the treatment of the two-electron repulsion integrals (ERIs), which constitute one of the primary memory bottlenecks in large-scale calculations~\cite{malone2018overcoming}. Most contemporary AFQMC implementations employ low-rank representations of the ERI tensor, such as the Cholesky decomposition, to reduce storage requirements~\cite{malone2018overcoming}.

Tensor hypercontraction (THC) methods~\cite{THC} based on interpolative separable density fitting (ISDF)~\cite{malone2018overcoming}, provide an alternative low-rank factorization of the ERI tensor that offers substantially improved compression. Originally introduced in quantum chemistry to reduce the storage and computational costs associated with four-index electron repulsion integrals, THC methods have recently been adapted to AFQMC~\cite{THC}, where Malone \textit{et al.} showed that the THC representation of the ERI tensor can dramatically reduce the memory requirements of AFQMC calculations while preserving chemical accuracy~\cite{malone2018overcoming}.}

In this work, we present a novel, more efficient approach to AFQMC that diagonalizes the quantum-chemical two-body interaction term in an extended space with additional fictitious modes. By diagonalizing, we mean that the two-body interaction term depends only on fermionic number operators. This simplification technique, also known as Isometric Tensor Hypercontraction (ITHC), was first proposed in Refs.~\cite{THC, Luo.2025} in the context of quantum computation, reducing the resources required to simulate the dynamics of chemical systems. When applied to AFQMC, the resulting Hubbard-like interaction enables the propagation of the AFQMC walkers with reduced computational complexity and memory usage. \changed{In contrast to the THC-AFQMC methods already reported in the literature~\cite{malone2018overcoming}, our method shows significant improvement in the runtime of both propagator and estimator, due to the diagonal form of the two-body term in the extended basis.} To date, novel methods that accelerate walker propagation have been scarce; thus, our proposed method represents a new approach to improving AFQMC methods in general~\cite{review_afqmc}.

\section{Theory}
\label{sec:theory}

\changed{This section explains our main algorithmic development: how to employ the isometric tensor hypercontraction form of the electron repulsion integral tensor in AFQMC calculations.}

\subsection{Notation}

\begin{table}[!ht]
\centering
\renewcommand{\arraystretch}{1.25}
\begin{tabular}{l|l}
    \textbf{Notation} & \textbf{Description} \\ \hline
    $N$ & Number of orbitals (original basis) \\
    $N_e$ & Number of electrons \\ 
    $N_\text{aux}$ & Number of orbitals (extended basis) \\ & Number of  auxiliary fields \\ 
    $N_w$ & Number of walkers \\
    $p, q,r,s$ & Indices for orbitals in original space \\ 
    $\alpha, \beta$ & Indices for orbitals in extended space \\ 
    $\gamma$ & Indices for auxiliary vectors \\ 
    $i, j$ & Indices for walkers \\ 
    \hline
    ITHC & Isometric tensor hypercontraction \\
    ISDF & Interpolative separable density fitting \\
    FCI& Full configuration interaction \\
    MBE-FCI & Many-body expanded FCI \\
    CCSDT & Coupled clusters singles doubles triples \\
    HS & Hubbard-Stratonovich \\
    \hline
\end{tabular}%
\caption{Notation and abbreviations used in this work.}
\label{Notation}
\end{table}
In Table~\ref{Notation}, we summarize the notation and abbreviations used in this paper. 

\subsection{Diagonalizing the interaction Hamiltonian}

The molecular electronic Hamiltonian can be written in second quantization using fermionic annihilation and creation operators as
\begin{equation}
\begin{split}
\hat{H} &= \hat{h} + \hat{V}
\\ &= \sum_{p,q}^{N} h_{pq} \hat{a}^{\dagger}_{p} \hat{a}_{q} + \frac{1}{2} \sum_{p,q,r,s}^N V_{pqrs} \hat{a}^{\dagger} _{p} \hat{a}^{\dagger}_{r} \hat{a}_{s} \hat{a}_{q}.
\end{split}
\end{equation}
Here, $\hat{a}^{\dagger}_{p}$ and $\hat{a}_{q}$ are the fermionic creation and annihilation operators, $h_{pq}$ the one-electron integrals and $V_{pqrs}$ the two-body integrals. Most AFQMC algorithms for quantum chemical problems rely on a Cholesky decomposition of the two-electron repulsion integral (ERI) tensor $V_{pqrs}$~\cite{Malone.2023, Lee.2020}. In our method, however, we employ the isometric tensor hypercontraction (ITHC) technique recently proposed in~\cite{Luo.2025}, which factorizes the ERI as 
\begin{equation}
    V_{pqrs} = \sum_{\alpha,\beta=1} ^{N_{\text{aux}}} u_{p\alpha} u_{q\alpha} W_{\alpha\beta} u_{r\beta} u_{s\beta},
\end{equation}
where $N_{\text{aux}}\geq N$ and $u_{i\alpha}$ is a real isometry. Unlike generic tensor hypercontraction \cite{hohenstein_tensor_2012,parrish_tensor_2012,lu_compression_2015,dong_interpolative_2018,lee_systematically_2020,matthews_improved_2020}, this special isometric form admits a simple physical interpretation.

Specifically, we introduce $N_{\text{aux}}-N$ additional fictitious modes with annihilation operators $b_m$, and define a new set of fermionic modes given by the annihilation operators via
\begin{equation}
    \hat{c}_\alpha = \sum_p u_{p\alpha} \hat{a}_p +\sum_m v_{m\alpha} \hat{b}_m.
    \label{eq:basis_rot}
\end{equation}
This is achieved via the canonical transformation given by $u$ and a complementary isometry $v$, namely, the concatenation of $u$ and $v$ is a square, orthogonal matrix. It allows us to rewrite the electron-electron interaction term into a Hubbard-type interaction in the extended space,
\begin{equation}
\label{eq:extended-two-body}
    \hat{W} = \frac{1}{2} \sum_{\alpha \neq \beta}^{N_{\text{aux}}}  W_{\alpha \beta} \hat{n}_{\alpha} \hat{n}_{\beta},
\end{equation}
where $\hat{n}_\alpha = \hat{c}^\dagger_\alpha \hat{c}_\alpha $ is the fermionic number operator in the extended space. The interaction operator given in Eq.~\eqref{eq:extended-two-body} recovers the original two-body term when projected back onto the physical Hilbert space:
\begin{equation}
    \hat{V} = \mathbf{P} \hat{W} \mathbf{P}.
\end{equation}
Here $\mathbf{P}$ is the projector to the physical Hilbert space with no fictitious modes. 

It is shown in~\cite{Luo.2025} that ITHC has a similar accuracy as generic THC for the same number of auxiliary modes $N_{\text{aux}}$. The authors also provided an efficient three-step method for obtaining ITHC, starting with a generic THC generated by the interpolative separable density fitting (ISDF) approach~\cite{lu_compression_2015} and transforming it into the required isometric form. We also employed the same method to obtain the ITHC factors in this study\changed{, for which a brief review is provided in Appendix \ref{app:metric}.}

\subsection{Extended basis AFQMC}
\label{subsec:extendedbasisAFQMC}

AFQMC is based on the imaginary time propagation~\cite{Review_of_QMC}
\begin{equation}
    \ket{\Psi_0} = \lim_{\tau \to \infty} e^{-\tau \hat{H}} \ket{\Psi_{\text{init}}} = \lim_{n \to \infty} \big(e^{-\Delta \tau \hat{H}}\big)^n \ket{\Psi_{\text{init}}},
    \label{eq:ground-state-projection}
\end{equation} 
where $\Delta \tau$ is an small time step, $\ket{\Psi_0}$ is the ground state and $\ket{\Psi_{\text{init}}}$ is an initial state not orthogonal to $\ket{\Psi_0}$. 

While the ITHC technique was originally proposed for unitary time propagation~\cite{Luo.2025}, it applies equally well to imaginary time propagation. For a short time step $\Delta \tau$, we make use of the second-order Trotter-Suzuki decomposition~\cite{Trotter.1959} and quantum Zeno dynamics~\cite{Facchi_2008,Burgarth2020quantumzenodynamics}, writing
\begin{equation}
    e^{-\Delta \tau \hat{H}} = e^{- \frac{\Delta \tau }{2} \hat{h}} \mathbf{P} e^{-\Delta \tau \hat{W}}\mathbf{P}  e^{- \frac{\Delta \tau }{2} \hat{h}}  + \mathcal{O}(\Delta \tau ^2) .
    \label{eq:Trotter-decomp}
\end{equation}
Notice that here the error is dominated by the Zeno error, while the Trotter error remains $\mathcal{O}(\Delta \tau ^3)$~\cite{Luo.2025}. A detailed analysis of different sources of error is provided in Appendix~\ref{app:errors}. This Zeno error translates to an error of order $\mathcal{O}(\Delta \tau)$ in the final equilibrium energy, as we can think of our simulation evolving under the effective Hamiltonian $\hat{H}_{\text{eff}} = \hat{H} + \mathcal{O}(\Delta \tau)$.

We realize the imaginary time step in Eq.~\eqref{eq:Trotter-decomp} as follows: after applying $e^{-\frac{\Delta \tau }{2} \hat{h}}$, we rotate the walkers to the extended basis, where the Monte Carlo calculation of the two-body interaction term $e^{-\Delta \tau \hat{W}}$ can be performed using solely number operators. We then rotate back and project onto the physical Hilbert space. For a Slater-determinant walker, these basis changes are equivalent to matrix multiplications with an isometry $u_{i\alpha}$.

The application of $e^{-\Delta \tau \hat{W}}$ resembles other AFQMC algorithms. We first write $\hat{W}$ as a sum of squares of one-body operators
\begin{equation}
\begin{split}
\label{eq: cholesky decomp of W}
    \hat{W} &= \frac{1}{2} \sum_{\alpha \neq \beta}^{N_{\text{aux}}} W_{\alpha \beta} \hat{n}_{\alpha} \hat{n}_{\beta} \\ &= \frac{1}{2} \sum_{\alpha \neq \beta}^{N_{\text{aux}}} \sum_{\gamma = 1}^{N_{\text{aux}}} w_{\gamma \alpha} w_{\gamma \beta} \hat{n}_{\alpha} \hat{n}_{\beta} = \frac{1}{2}\sum_{\gamma = 1}^{N_{\text{aux}}} \hat{w}_{\gamma}^2.
\end{split}
\end{equation}
Here, the matrix $w_{\gamma \alpha}$ can be obtained via an eigen- or Cholesky decomposition of $W_{\alpha\beta}$, which is a positive semidefinite matrix. After that, a Hubbard-Stratonovich transformation~\cite{Hubbard.1959} is performed to yield the AFQMC propagator given as
\begin{align}
\label{eq:AFQMC-two-body-propagator}
    e^{-\Delta \tau \hat{H}} &= \int d\mathbf{x}^{N_{\text{aux}}} p(\mathbf{x}) \hat{B}(\mathbf{x}) + O(\Delta \tau ^2), \\
    \hat{B}(\mathbf{x}) &= e^{- \frac{\Delta \tau}{2}\hat{h} } \mathbf{P} e^{\sqrt{-\Delta \tau} \mathbf{x} \cdot \hat{\mathbf{w}}} \mathbf{P} e^{-\frac{\Delta \tau}{2} \hat{h}},
\label{eq:extended-propagator}
\end{align}
where $\mathbf{x}$ is an auxiliary field vector with ${N_{\text{aux}}}$ entries and $p(\mathbf{x})$ denotes the normal distribution. Note that  $\hat{B}(\mathbf{x})$ is a one-body operator.

The transformation given in Eq.~\eqref{eq:AFQMC-two-body-propagator} maps an interacting many-body problem onto a sum of non-interacting one-body problems, where the fluctuation beyond the mean-field is recovered by evaluating the integral over all auxiliary field configurations~\cite{Hubbard.1959}. The integral in Eq.~\eqref{eq:AFQMC-two-body-propagator} is evaluated by sampling the auxiliary field $\mathbf{x}$ from the probability distribution $p(\mathbf{x})$ using Monte Carlo methods.
Using this propagator, in principle, one may evolve a set of walkers to realize the ground-state projection scheme from Eq.~\eqref{eq:ground-state-projection}. Free-projection AFQMC (fp-AFQMC), however, suffers from the well-known fermionic sign or phase problem found in fermionic QMC methods~\cite{Zhang.2003}. Due to the inherently anti-symmetric nature of fermionic wave functions, the fermionic sign problem of QMC~\cite{Sign_problem, Zhang.2003} will cause contributions with different signs to cancel each other. This leads to an exponential decay of the signal-to-noise ratio in the QMC sampling process~\cite{Zhang.2003}.

To overcome the fermionic phase problem in AFQMC, the phaseless approximation (ph-AFQMC) is used, which employs importance sampling to bias the sampling process and fix walkers' phases~\cite{Motta.2018, Zhang.2003}. In ph-AFQMC, the wave function is represented as a weighted sum over a set of walkers $\{  \ket{\phi_i(\tau)} \}$ with weights $w_i(\tau)$
\begin{equation}
    \ket{\Psi(\tau)} = \sum_{i=1}^{N_{w}} w_i(\tau) \frac{\ket{\phi_{i}(\tau)}} {\braket{\Psi_{\text{trial}} | \phi_{i}(\tau)}}.
\end{equation}
Here, $\ket{\Psi_{\text{trial}}}$ is the trial wavefunction employed for the importance sampling. Then, the global energy estimator at a given imaginary time $\tau$ is given by the weighted statistical average of local energies
\begin{equation}
    E(\tau) = \sum_i^{N_{w}} w_i(\tau) E_{\text{loc},i} (\tau) = \sum_i^{N_{w}} w_i(\tau)\frac{\bra{\Psi_{\text{trial}}} \hat{H} \ket{\phi_i(\tau)}}{\langle \Psi_{\text{trial}}| \phi_i(\tau) \rangle}. \label{eq:localenergy}
\end{equation}
The walker state $\ket{\phi_i (\tau)}$ and the weight $w_i(\tau)$ are updated by the following rules
\begin{align}
    \ket{\phi_i(\tau + \Delta \tau)} &= \hat{B}(\mathbf{x}_i-\bar{\mathbf{x}}_i) \ket{\phi_i (\tau)}, \label{eq:propagation}
    \\ w_i (\tau + \Delta\tau) &= I_{\text{ph}}(\mathbf{x}_i, \bar{\mathbf{x}}_i,\tau,\Delta \tau) \times w_i (\tau).
    \label{eq:update_weight}
\end{align}
Note the shift in the integral variable $\mathbf{x}$ in Eq.~\eqref{eq:AFQMC-two-body-propagator} by the force bias $\bar{\mathbf{x}}$. It can be shown~\cite{Zhang.2003} that the optimal choice for $\bar{\mathbf{x}}$ is given by
\begin{align}
    \bar{\mathbf{x}}_i = - \sqrt{\Delta\tau} \frac{\bra{\Psi_{\text{trial}}}\hat{\mathbf{w}} \ket{\phi_i} }{\braket{\Psi_{\text{trial}}|\phi_{i}}}. \label{eq:forcebias}
\end{align}
The corresponding phaseless importance function $I_{\text{ph}}$ is given by
\begin{equation}
    I_{\text{ph}} (\mathbf{x}_i, \bar{\mathbf{x}}_i,\tau,\Delta \tau) = \abs{ S_i (\tau, \Delta \tau) e^{\mathbf{x}_i \cdot \bar{\mathbf{x}}_i - \bar{\mathbf{x}}_i \cdot \bar{\mathbf{x}}_i/2 }} 
    \times \max\left(0, \cos(\arg(S_i (\tau, \Delta \tau)))\right). 
\end{equation}
Here $S_i (\tau,\Delta \tau)$ is the overlap ratio of the $i$-th walker
\begin{equation}
    S_i(\tau,\Delta \tau) = \frac{\bra{\Psi_{\text{trial}}} \hat{B}(\mathbf{x}_i - \bar{\mathbf{x}}_i) \ket{\phi_i (\tau)}}{ \braket{\Psi_{\text{trial}} | \phi_i (\tau)}}.
\end{equation}
This modified update rule ensures that the weights $w_i$ remain real and positive throughout propagation but introduces a systematic bias, which vanishes if the trial state $\ket{\Psi_{\text{trial}}}$ is exactly the ground state. Finally, we noted that the mean-field shift with respect to the trial wavefunction $\ket{\Psi_{\text{trial}}}$ can be employed to further reduce the variance $w_i$~\cite{jiang2024improved}.

The propagation of walkers in Eq.~\eqref{eq:propagation}, the evaluation of the force bias in Eq.~\eqref{eq:forcebias}, and the evaluation of local energies in Eq.~\eqref{eq:localenergy} are computationally costly and need special treatment in our extended basis AFQMC framework. We explain the implementation details of these parts in Sec.~\ref{sec:implementation}.

\section{Implementation Details}
\label{sec:implementation}

In this section, we describe the implementation details of our extended basis AFQMC method. To realize our extended basis AFQMC propagator, we utilized the Python package ipie by Joonho Lee \textit{et al.}~\cite{Malone.2023, jiang2024improved} as a basis and added an additional propagator and other functionalities.

\subsection{Enlarged basis propagator}

Since we are using single Slater-determinant states, the application of the imaginary-time propagator in Eq.~\eqref{eq:extended-propagator} can be performed as a simple matrix-matrix multiplication according to Thouless theorem~\cite{Thouless.1960}. 

After applying the one-body propagator $e^{-\Delta\tau \hat{h}/2}$ in the original basis with dimension $N$, the $i$-th walker which is represented as a determinant matrix $B^{(i)}$ with dimension $(N,N_e)$, is rotated to the extended space via
\begin{equation}
\label{eq:half-rotation}
    \tilde{B}^{(i)}_{\alpha k} = \sum_{p=1}^{N} u_{p\alpha} B^{(i)}_{pk}.
\end{equation}
Here, $u_{p \alpha}$ is the basis rotation matrix in Eq.~\eqref{eq:basis_rot}. The rotation of the Slater determinants, as stated in Eq.~\eqref{eq:half-rotation}, can be performed at a cost of $\mathcal{O}(N_{w} N_{\text{aux}} N N_e)$. 

We then apply the Hubbard-Stratonovich transformed two-body propagator in the extended basis as
\begin{align}
    (e^{ \sqrt{- \Delta \tau} \mathbf{x}\cdot \mathbf{w}})_{\alpha\alpha} \tilde{B}_{\alpha k}.
\end{align}
Since the two-body interaction term is diagonal in the extended space, i.e., only contains terms proportional to the number operators $\hat{n}_\alpha$, this can be done at a cost of $\mathcal{O}(N_{w}N_{\text{aux}}N_e)$.

\subsection{One-body Green's function and force bias}
\label{sec:One-body Green's function and force bias}

An important component of the ph-AFQMC is the calculation of the force bias~\cite{Malone.2023, Zhang.2003}. In our setup, the force bias is calculated in the extended basis by
\begin{align}
    \bar{x}_{i,\gamma} = - \sum_\alpha v_{\gamma \alpha}  \sqrt{\Delta\tau} \frac{\bra{\Psi_{\text{trial}}}\hat{n}_\alpha \ket{\phi_i} }{\braket{\Psi_{\text{trial}}|\phi_{i}}}.
\end{align}
Therefore, the mixed estimation of $\hat{n}_\alpha$ is needed, which is the diagonal part of the Green's function in the extended basis.

The one-body Green's function can be computed efficiently by first performing a rotation of the trial state and walkers to the enlarged basis, as in Eq.~\eqref{eq:half-rotation}, obtaining the rotated determinant matrices $\tilde{A}$ and $\tilde{B}$ for the trial and walker determinants, respectively. Notice that the rotation of the trial state matrix $A$ needs to be performed only once at the beginning of the simulation. Then the Green's function in the extended basis is calculated using
\begin{equation}
\tilde{G} = \tilde{A}^* (\tilde{B}^T \tilde{A})^{-1} \tilde{B}^T  =  \tilde{A}^* (B^T A)^{-1} \tilde{B}^T,
\end{equation}
where we have used the fact that $u$ is an isometry. 
The evaluation of the full Green's function in the extended basis has a theoretical complexity of $\mathcal{O}(N_{w} N_{\text{aux}}^2 N_e)$. As only the diagonal elements of $\tilde{G}$ are required to compute the force bias, the cost can be further reduced to $\mathcal{O}(N_{w}N N_e^2 + N_{w} N_{\text{aux}} N_e )$. Then the force bias is computed via a simple matrix vector product, with cost $\mathcal{O}(N_{w} N_{\text{aux}}^2)$.

The pseudocode for a complete implementation of one propagation step is given in Alg.~\ref{alg:propagation}. The dominant computational cost arises from the basis transformation, leading to an overall complexity of $\mathcal{O}(N_w N_{\mathrm{aux}} N N_e )$, with a memory usage of $\mathcal{O}(N_{\mathrm{aux}}^2)$. 
For comparison, the complexity of the standard AFQMC propagator based on Cholesky decomposition or density fitting~\cite{Malone.2023} is $\mathcal{O}(N_w N_{\mathrm{aux}} N^2)$ and the memory usage is $\mathcal{O}(N_{\mathrm{aux}} N^2)$, since it requires the handling of large Cholesky tensors. It is known that to achieve the same accuracy, the rank of the Cholesky tensor $N_{\text{aux}}= N_{\mathrm{chol}}$ is usually smaller than the rank of THC decomposition $N_{\text{aux}}=N_{\mathrm{ithc}}$. Therefore, the complexity comparison does not necessarily imply a speedup in a realistic implementation. However, in our numerical experiments, we observe reduced runtimes on GPUs for the linear hydrogen chain benchmark, as shown in Sec.~\ref{sec:Timing Benchmarks}. 

\begin{figure}[t]
\centering
\begin{minipage}{0.95\columnwidth}
\hrule
\vspace{0.5em}
\textbf{Algorithm 1: Imaginary time propagation}
\vspace{0.5em}
\begin{algorithmic}[1]
\State \textbf{Input:} Initial batch of walkers $\{ \ket{\phi_{i}} \}$ as Slater determinants, and weights $w_i$
\State \textbf{Output:} Propagated walkers and updated weights
\For{each walker $i = 1, \dots, N_{w}$}
    \State Compute overlap: $ \braket{\Psi_{\text{trial}} | \phi_i} \to \text{ovlp}$
    \State Apply one-body propagator: $ e^{-\Delta \tau \hat{h}/2} \ket{\phi_i} \to \ket{\phi_i}$
    \State Rotate the walkers: $u \ket{\phi_i} \to \ket{\tilde{\phi}_i}$
    \State Sample auxiliary fields $\mathbf{x}_i$ from $p(\mathbf{x})$
    \State Compute the diagonal part of Green's function $\expval{\hat{\mathbf{n}}_i}$
    \State Compute force bias: $-\sqrt{\Delta \tau} w \expval{\hat{\mathbf{n}}_i} \to \bar{\mathbf{x}}_i$
    \State Force bias factor: $\mathbf{x}_i \cdot \bar{\mathbf{x}}_i - \bar{\mathbf{x}}_i \cdot \bar{\mathbf{x}}_i /2 \to c_{\text{fb}}$
    \State Two-body propagator: $\sqrt{-\Delta \tau} (\mathbf{x}_i - \bar{\mathbf{x}}_i) \cdot \hat{\mathbf{w}} \to \hat{T}$
    \State Apply two-body propagator: $e^{\hat{T}} \ket{\tilde{\phi}_i} \to \ket{\tilde{\phi}_i}$
    \State Rotate back to original space: $u^{-1} \ket{\tilde{\phi}_i} \to \ket{\phi_i}$
    \State Apply one-body propagator: $e^{-\Delta \tau \hat{h}/2} \ket{\phi_i} \to \ket{\phi_i}$
    \State Compute overlap: $\braket{\Psi_{\text{trial}} | \psi_i} \to \text{ovlp}_{\text{new}}$
    \State Update weight $w_i$ according to Eq.~\eqref{eq:update_weight}
\EndFor
\end{algorithmic}
\vspace{0.5em}
\hrule
\end{minipage}
\caption{Imaginary-time propagation procedure for the extended-basis AFQMC method.}
\label{alg:propagation}
\end{figure}

\subsection{Energy estimator}

The local energy is determined via the mixed estimator~\cite{review_afqmc}
\begin{equation}
    E_{\text{loc},i} = \frac{\braket{\Psi_{\text{trial}} | \hat{H} | \phi_i}}{\braket{\Psi_{\text{trial}} | \phi_i}}.
\end{equation}
For Slater-determinant states, this can be computed via the one-particle Green's function \cite{Zhang.2003}. While the one-body energy is evaluated in the original space, we evaluate the two-body interaction energy in the extended space 
\begin{equation}
\begin{split}
    \expval*{\hat{V}} &= \frac{1}{2}\sum_{\alpha \neq \beta} W_{\alpha\beta} \expval{\hat{n}_{\alpha} \hat{n}_{\beta}} \\
    &= \frac{1}{2} \sum_{\alpha \neq \beta} W_{\alpha \beta} (\tilde{G}_{\alpha \alpha} \tilde{G}_{\beta \beta}- \tilde{G}_{\alpha \beta} \tilde{G}_{\beta \alpha}),
\end{split}
\end{equation}
where $\tilde{G}_{\alpha \beta}$ is the Green's function in the extended basis and $W_{\alpha \beta}$ is the transformed ERI in the extended basis representation.

The computational cost of evaluating the global energy is dominated by the computation of the Green's function in the extended basis, resulting in a computational complexity of $\mathcal{O}(N_w N_{\mathrm{aux}}^2 N_e)$ and a memory requirement of $\mathcal{O}(N_{\mathrm{aux}}^2)$. In contrast, the standard ph-AFQMC method costs $\mathcal{O}(N_w N_{\mathrm{aux}} N N_e^2)$ for energy estimation~\cite{Malone.2023}. Our method thus provides a significant reduction in computational cost. Note that previous studies used generic THC to accelerate energy evaluation~\cite{malone2018overcoming}, which exhibits computational and memory scaling similar to our method.

\subsection{Algorithmic scaling}

\changed{In this section, we summarize the algorithmic scaling of our ITHC-AFQMC method in terms of complexity and memory usage, and compare the results to existing THC methods found in the literature. 

Previous applications of the THC (based on ISDF) in AFQMC have been reported in the literature by  Malone \textit{et al.}~\cite{malone2018overcoming}, where they reduced the memory usage down to $\mathcal{O}(N_\text{aux}^2)$~\cite{malone2018overcoming} and energy estimator complexity to $\mathcal{O}(N_\text{aux}^2N_e)$. However, in this framework, THC primarily serves as a compression technique and retains the conventional structure of the AFQMC two-body propagator, which is why the authors reported no significant practical speed-up compared to standard AFQMC~\cite{malone2018overcoming}. 

In contrast, the key distinction of ITHC is that it serves not only as a mathematical compression technique but also admits a physical interpretation. In ITHC, the two-body interaction becomes diagonal in the extended basis. This diagonal form significantly simplifies the propagator and local energy estimator. As a consequence, the computational hotspots of the AFQMC workflow, namely walker propagation and local energy evaluation, can be rewritten in terms of diagonal interactions. Because of the reasons listed above, the ITHC method not only acts as a compression method but also significantly speeds up key computational steps, something the THC-AFQMC method mentioned earlier was unable to do~\cite{malone2018overcoming}.}

\begin{table}[h]
    \centering
    \begin{tabular}{c|c|c|c}
       Method  & Memory & Propagator & Energy Estimator  \\
       \hline
        Standard~\cite{Malone.2023} & $\mathcal{O}(N_\text{aux}N^2)$ & $\mathcal{O}(N_\text{aux}N^2)$ & $\mathcal{O}(N_\text{aux}NN_e^2)$\\
        THC \cite{malone2018overcoming} &$\mathcal{O}(N_\text{aux}^2)$& & $\mathcal{O}(N_\text{aux}^2N_e)$\\
        ITHC (ours) &$\mathcal{O}(N_\text{aux}^2)$& $\mathcal{O}(N_\text{aux}NN_e)$ & $\mathcal{O}(N_\text{aux}^2N_e)$
    \end{tabular}
    \caption{Comparison of memory performance, propagator complexity, and energy estimator complexity across different AFQMC methods. $N$ denotes the number of orbitals in the original basis, $N_e$ the number of electrons, and $N_\text{aux}$ the the number of auxiliary fields. The THC-AFQMC paper\cite{malone2018overcoming} does not provide a detailed discussion of the propagator complexity, but reports that no significant speedup was observed in numerical experiments.}
    \label{tab:Memory comparison}
\end{table}
\changed{We summarized the above discussion in Table~\ref{tab:Memory comparison}. The improvements in terms of complexity and memory are further supported by the benchmarking experiments of the $H_{10}$-chain, which are discussed in Section~\ref{sec:Timing Benchmarks}.}

\section{Applications and Benchmarking}
\label{sec:application}

In this section, we demonstrate that our extended basis AFQMC method is a viable addition to the ipie package by applying it to two quantum-chemical systems: the hydrogen chain toy model and the benzene molecule.
First, we benchmark the accuracy of our algorithm by computing the ground-state energy of the $\ce{H10}$ chain.
Furthermore, we benchmark the time scaling of our method against the hydrogen chain with various chain lengths. \changed{We discuss and compare the results to the standard AFQMC approach. Additionally, we implement an intermediate method that only uses the ITHC method to compress the ERI (without rotating the basis) as an additional reference. Details of the intermediate method are discussed in Appendix~\ref{app:Intermediate}.}
Finally, we demonstrate the algorithm's ability to capture electronic correlation by computing the correlation energy of benzene beyond the mean-field approximation\changed{, and analyze the convergence behavior of the correlation energy with respect to the ITHC rank.} 
All calculations were run on a single NVIDIA Tesla V100 GPU.

\changed{In Appendix~\ref{app:metric}, we briefly review the ITHC pre-calculation procedure and report its wall time. We also provide the ranks and approximation errors of the ITHC factorization used in extended AFQMC and the Cholesky decomposition employed in standard AFQMC. }
The ITHC data and the integrals for the systems studied in this work are openly available in Ref.~\cite{zenodo_19368252}.

\subsection{$\ce{H10}$-chain ground state energy}
\label{sec: H10 chain}
To confirm the accuracy and efficiency of our algorithm, multiple benchmarks were run using the $\ce{H10}$-chain toy model with an STO-6G minimal basis set and a single-determinant trial wavefunction derived from an unrestricted Hartree-Fock (UHF) calculation~\cite{Slater.1951}. The interatomic distance was set to 1.6~bohr for the hydrogen chain. 
An AFQMC simulation was run using the standard method in ipie and our extended method.  
\changed{Additionally, to demonstrate that the advantage of our method arises from the enlarged-basis framework rather than the ITHC decomposition itself, we performed an "intermediate method" simulation. In this approach, the ITHC is used to generate the Cholesky vectors, as discussed in Appendix~\ref{app:Intermediate}, which are then used as inputs to the standard ipie workflow.}

\begin{figure}
    \centering
    \includegraphics[scale=0.6]{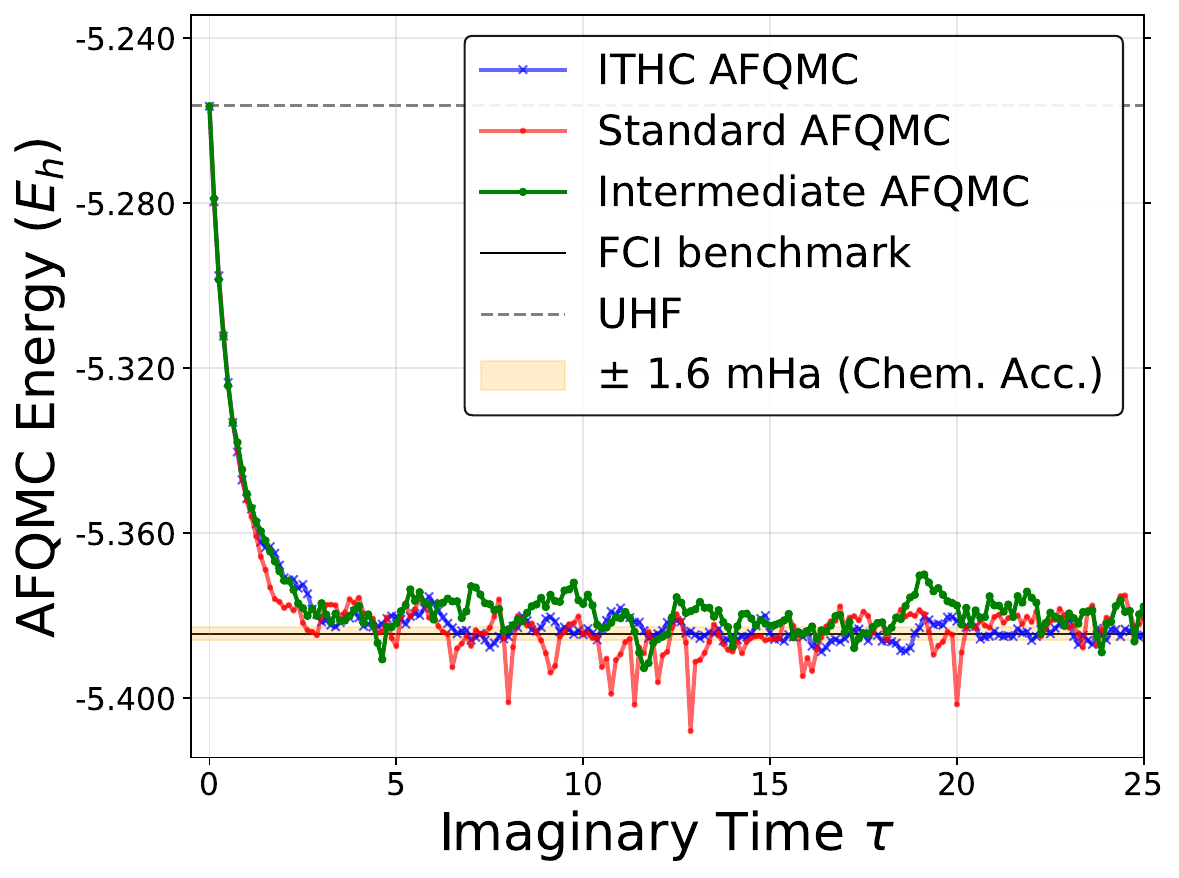}
    \caption{Convergence of the AFQMC simulation energy of the $\ce{H10}$-chain using the standard AFQMC method (red), our ITHC method (blue), and the intermediate method (green). The HF ground-state energy is shown as a black dotted line. The FCI benchmark is shown as a black line with the shaded area representing the range of chemical accuracy ($\pm 1.6~mE_h$).}
    \label{fig:H10 chain convergence}
\end{figure}

The AFQMC simulations are run with 1000 walkers, \changed{a time step of $\Delta \tau = 0.005$~a.u., and a total evolution time of $25$~a.u.} The local energy is estimated once per block, which contains $10$ imaginary time steps to minimize computational effort while maintaining a significant sample size. \changed{The results of the three different AFQMC methods are shown in Fig.~\ref{fig:H10 chain convergence}.}

The progression of the AFQMC simulation is shown in Fig.~\ref{fig:H10 chain convergence}. \changed{After an initial equilibration period, the AFQMC simulations converged to an average value shown by the blue line (our method), the red line (standard method), and the green line (intermediate method).} By discarding the first data points before \changed{$\tau  = 6.25$ ~a.u.} and taking the average, an estimate of the ground-state energy can be obtained. \changed{Notably, our extended ITHC AFQMC approach maintains significantly lower stochastic noise and fluctuations than the standard and intermediate methods.} 

\begin{figure}[htbp]
\centering
\begin{tabular}{l|c|c|c}
& ITHC & Standard & Intermediate\\
\hline
Seed 1 & -5.383847 & -5.380919 & -5.382167\\
Seed 2 & -5.383656 & -5.381657 & -5.381872\\
Seed 3 & -5.383896 & -5.384778 & -5.379899\\
Seed 4 & -5.384092 & -5.381426 & -5.384150\\
Seed 5 & -5.382526 & -5.381946 & -5.379865\\
\hline
Avg. & -5.383603 & -5.382145 & -5.381591\\
Std. dev. & 0.000556 & 0.001358 & 0.001600\\
Error & 0.000757 & 0.002215 & 0.002770\\
\end{tabular}

\vspace{0.8cm}

\includegraphics[scale=0.5]{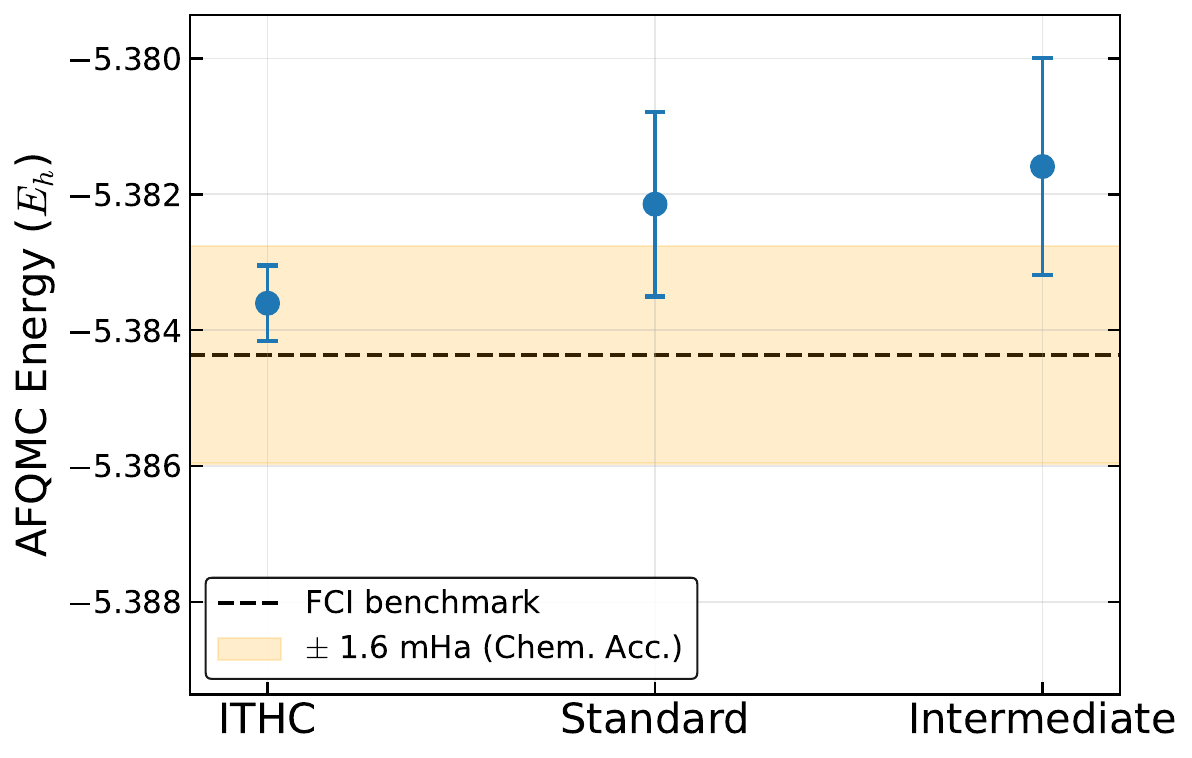}
\caption{
The table summarizes the converged energies obtained across five independent random seeds for different AFQMC methods, along with their means, standard deviations, and errors w.r.t. the FCI benchmark $-5.384361 E_h$. The lower panel visualizes the averaged AFQMC energies, with the error bars representing the standard deviation. The FCI benchmark is given as a black dotted line, and the shaded area represents the range of chemical accuracy ($\pm 1.6~mE_h$). 
}
\label{fig:H10_seeds}
\end{figure}

\changed{The results shown in Fig.~\ref{fig:H10 chain convergence} are based on a single run with one random seed. In practice, we find that results can vary substantially across random seeds. To quantify these statistical fluctuations and compare the numerical stability of the different methods, we performed five independent simulations for each method with different random seeds and report the corresponding mean values and standard deviations in Table~\ref{fig:H10_seeds}. Compared with the standard and intermediate methods, our ITHC approach yields results within chemical accuracy of the FCI benchmark $-5.384361 E_h$ and maintains substantially smaller standard deviations across the different tracks, as shown in Table~\ref{fig:H10_seeds} and visualized in Fig.~\ref{fig:H10_seeds}. By contrast, the intermediate method performs similarly to the standard method. We report a mean value of $-5.382145 E_h$ for the standard method, which is in line with previous results reported by Malone \textit{et al.}~\cite{Malone.2023}. These results indicate that our ITHC-AFQMC framework provides improved accuracy and numerical stability for the ground-state energy compared to the standard method, at least in the case of $H_{10}$. And these advantages arise from our unique propagation algorithm rather than from ITHC compression alone.}

\subsection{Timing benchmarks}
\label{sec:Timing Benchmarks}

It is well documented in the literature that the two most computationally demanding steps of AFQMC are propagation and local energy evaluation~\cite{Malone.2023}. In this section, we compare the computational efficiency of our proposed methodology with the standard AFQMC implementation in the ipie package. We demonstrate that our approach yields superior GPU performance in both the propagation and estimator phases, primarily due to the reduced theoretical complexity and reduced memory usage afforded by the Hubbard-like representation of the interaction in the extended space, as stated in Sec.~\ref {sec:One-body Green's function and force bias}. To quantify these gains, we benchmarked GPU runtimes for a series of one-dimensional hydrogen chains with variable length $N$, enabling us to evaluate performance as a function of system size.
\begin{figure}
    \centering
    \includegraphics[scale=0.6]{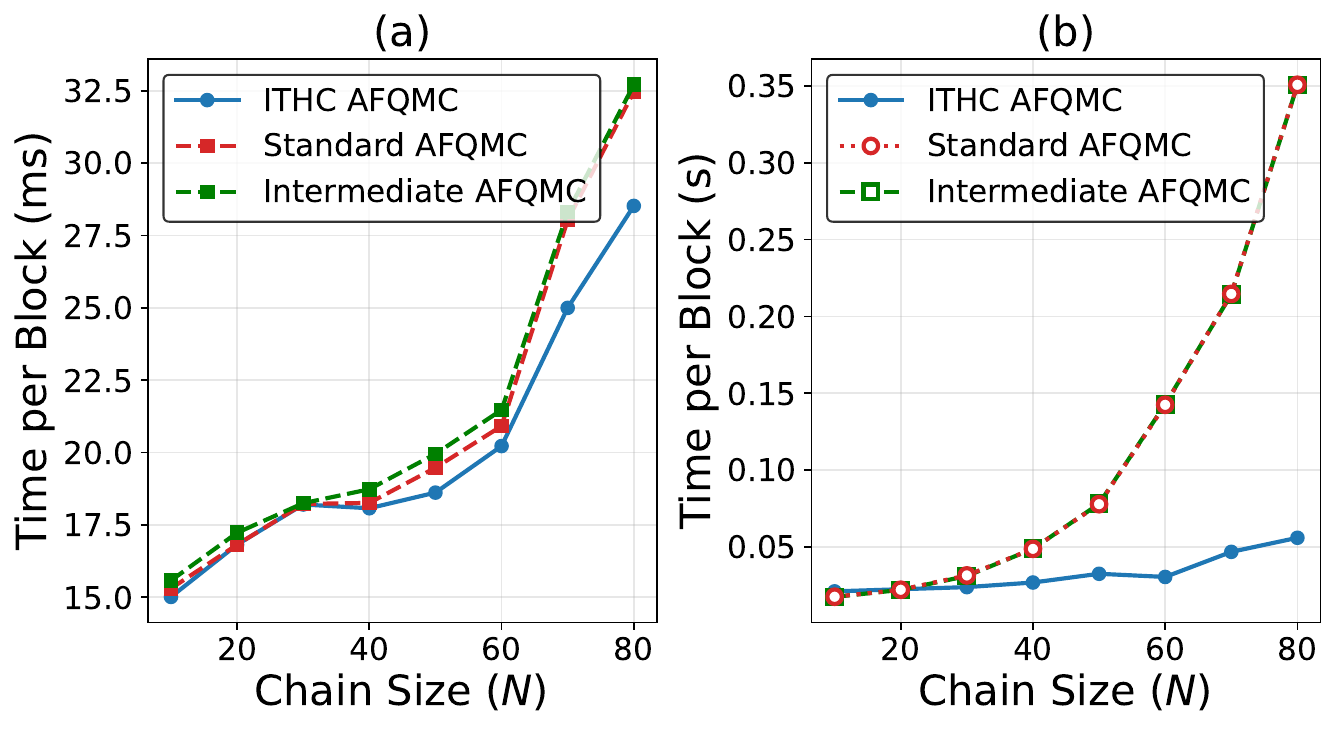}
    \caption{Comparison of GPU performance of propagation time (left) and estimator time (right) between our method (blue), intermediate method (green), and the standard method found in ipie (red) for different lengths of the hydrogen chain.}
    \label{fig:timing benchmark}
\end{figure}
\noindent As illustrated in Fig.~\ref{fig:timing benchmark}, the propagation times for both methods are comparable for smaller systems  ($N= 10-40$). However, as the chain size increases, our extended method exhibits better time performance. 

Regarding the estimator's performance, the standard method performs similarly to our method for chain sizes below $ N=20$. However, the results in Fig.~\ref{fig:timing benchmark} show that the estimator's runtime grows more rapidly for the ipie method. For larger systems, the estimator implemented in our method is significantly faster.

\changed{Timing benchmarks of the intermediate method show wall times and computational scaling identical to those of uncompressed Cholesky AFQMC within ipie, when performed at the same Cholesky rank shown in Fig.~\ref{fig:timing benchmark}. These results indicate that the observed speed-up is not a direct consequence of the ITHC decomposition itself, but is instead mainly attributable to the simplified form of the propagator in the extended basis.
Further discussion of the intermediate method is found in Appendix~\ref{app:Intermediate}.}

To summarize, the results in Fig.~\ref{fig:timing benchmark} demonstrate that our method achieves superior GPU performance in both propagation and estimator time, thanks to the lower time and memory complexity introduced by the Hubbard-like interaction term in the extended space. \changed{Benchmarking trials involving an intermediate implementation show that the reported speed-up is mainly attributed to the simplified form of the propagator.} This lower numerical complexity and more favorable scaling allow the application to larger molecular systems.

\subsection{Correlation energy of benzene and Zeno error}
\label{sec:Correlation Energy of Benzene and Zeno Error}

To determine our method's ability to capture electronic correlation in simple organic molecules, we calculated the correlation energy ($E_{\text{corr}}$) of benzene. The correlation energy is defined as~\cite{Szabo.2012}
\begin{align*}
    E_{\text{corr}}= E_{\text{AFQMC}}- E_{\text{UHF}}  .
\end{align*}
\noindent An AFQMC calculation was performed using our extended method and the standard method in the cc-pVDZ basis set, with a single-determinant UHF trial wavefunction and an active space of 30 electrons, using 1000 walkers.

To quantify the effects of finite-step errors, we calculated the average correlation energy of benzene for different step sizes, $\Delta \tau$. We performed extended AFQMC with time steps ranging from $\Delta \tau= 0.001$~a.u. to $\Delta \tau= 0.005$~a.u. for a total evolution time of $10$~a.u. By averaging the data points from $\tau=5$~a.u.\ onwards, an average value was obtained. To further reduce noise from statistical fluctuations, five independent simulations were run at each time step with different random seeds, and the results were averaged.
\begin{figure}
    \centering
    \includegraphics[width=0.6\linewidth]{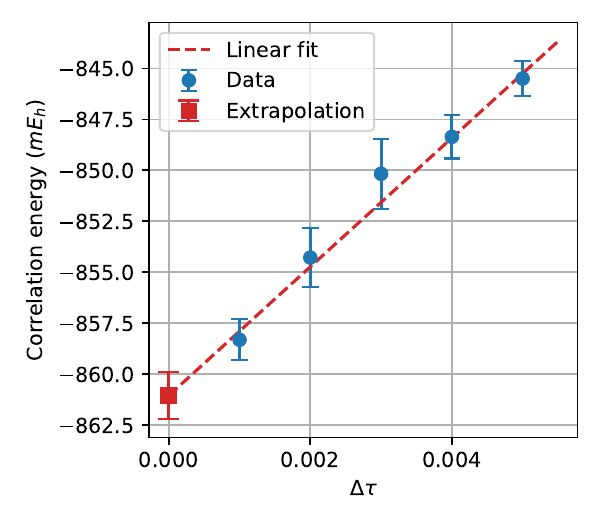}
    \caption{Average value of the correlation energy of benzene calculated using our extended method with different time steps $\Delta \tau$. The error bars indicate the standard deviation, and the linear regression fit is shown as a red dotted line. The extrapolated value for $\Delta \tau=0$ is $-861(1) mE_h$.}
    \label{Zeno error analysis}
\end{figure}

As shown in Fig.~\ref{Zeno error analysis}, the Zeno error becomes significant for large systems, which is consistent with the literature~\cite{Luo.2025}. The correlation energy exhibits a linear dependence on the time step $\Delta \tau$, agreeing with the theoretical prediction in Sec.~\ref{subsec:extendedbasisAFQMC}. Therefore, we performed a linear regression and extrapolated to the limit where $\Delta \tau = 0$, obtaining the predicted value in the absence of Zeno error. The accuracy of this extrapolated value is benchmarked against several high-level methods in Table~\ref{tab:results benzene}~\cite{Lee.2020}.
\begin{table}
    \centering
    \renewcommand{\arraystretch}{1.25}
    \begin{tabular}{c|@{\hspace{2mm}}l}
        Method             & $E_{\text{corr}}$ $(mE_h)$ \\
        \hline
         CCSDT             & $-859.9$ \\
         Extended AFQMC    & $-861(1)$ \\
         DMRG              & $-862.8(7)$ \\
         MBE-FCI           & $-863.0$ \\
         ph-AFQMC+CAS(6,6) & $-864.3(4)$ \\
         Standard ph-AFQMC & $-866.1(3)$ \\
    \end{tabular}
    \caption{The correlation energy ($E_\text{corr}$) of benzene in the cc-pVDZ basis set calculated via different methods \cite{Lee.2020}. The standard ph-AFQMC was run using an SD trial state, while the ph-AFQMC+CAS result was obtained via MSD trial states~\cite{Lee.2020}. Both ph-AFQMC simulations were performed using the QMCPACK package~\cite {Kim.2018}.}
    \label{tab:results benzene}
\end{table}
As summarized in Table~\ref{tab:results benzene} and Fig.~\ref{Zeno error analysis}, the extrapolation yields a predicted value of $-861(1) mE_h$, which is within $2 mE_h$ of the exact value reported in the literature~\cite{Lee.2020}. Notably, this amounts to a relative error of $0.2\%$ with respect to the many-body expanded
full configuration interaction (MBE-FCI) benchmark of $-863.0 mE_h$, which is agreed to be exact in the literature~\cite{Lee.2020}. While our method still underestimates the correlation energy relative to the exact value in the absence of Zeno error, the results are nonetheless more accurate than CCSDT. Given the significantly lower computational cost of our AFQMC method $\mathcal{O}(N^{3})$ compared to CCSDT $\mathcal{O}(N^{7})$ and MBE-FCI, these results demonstrate the high efficiency and potential of the extended-basis AFQMC method in describing many-body effects~\cite{Chemistry_review}.

Deviations from the exact value~\cite{Lee.2020} in the absence of Zeno errors likely occur due to the limitations of single Slater-determinant trial wavefunctions. Since single Slater-determinants inherently lack the multi-configurational character required to describe static correlation, a more appropriate choice is to use multi-determinant trial states. The authors of ipie reported a significant improvement in correlation energies when using a multi-determinant trial states generated from an initial Complete Active Space Self-Consistent Field (CASSCF)~\cite{Lee.2020}. This, however, will significantly increase the computational effort required to run an AFQMC simulation. Even though large multi-determinant wavefunctions may yield results that are close to exact, their unfavorable scaling is a costly trade-off~\cite{Lee.2020}.

\subsection{AFQMC performance with various ITHC ranks}
\label{sec:AFQMC performance with various ITHC ranks}
\begin{figure}[htbp]
    \centering
    \includegraphics[scale=0.6]{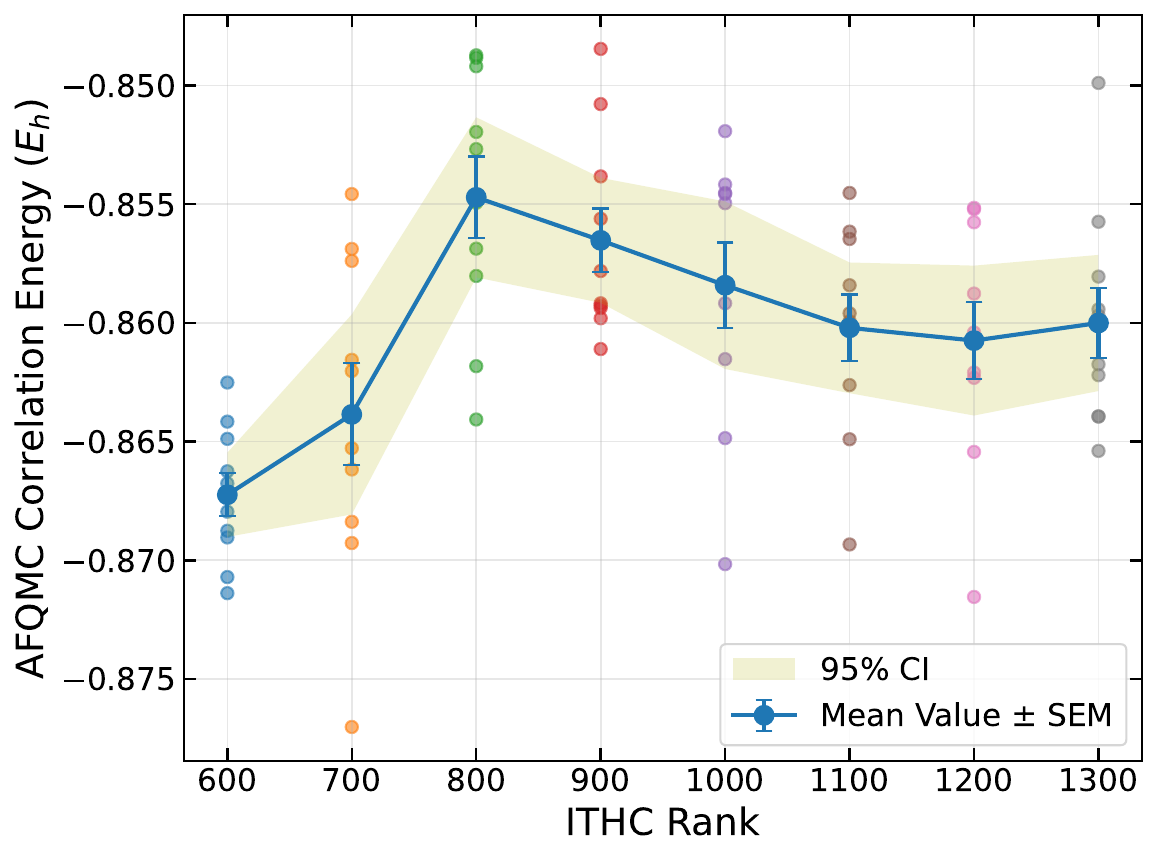}
    \caption{Converged AFQMC correlation energies of benzene for different ITHC ranks and a timestep $\Delta \tau= 0.001$. Points of the same color indicate the values for different random seeds. The large blue points are averages over all seeds. Different ITHC ranks are shown in different colors. The error bars show the Standard Error of Means (SEM), while the shaded area shows the 95 \% confidence interval.}
    \label{fig:ITHC_AFQMC}
    \includegraphics[scale=0.5]{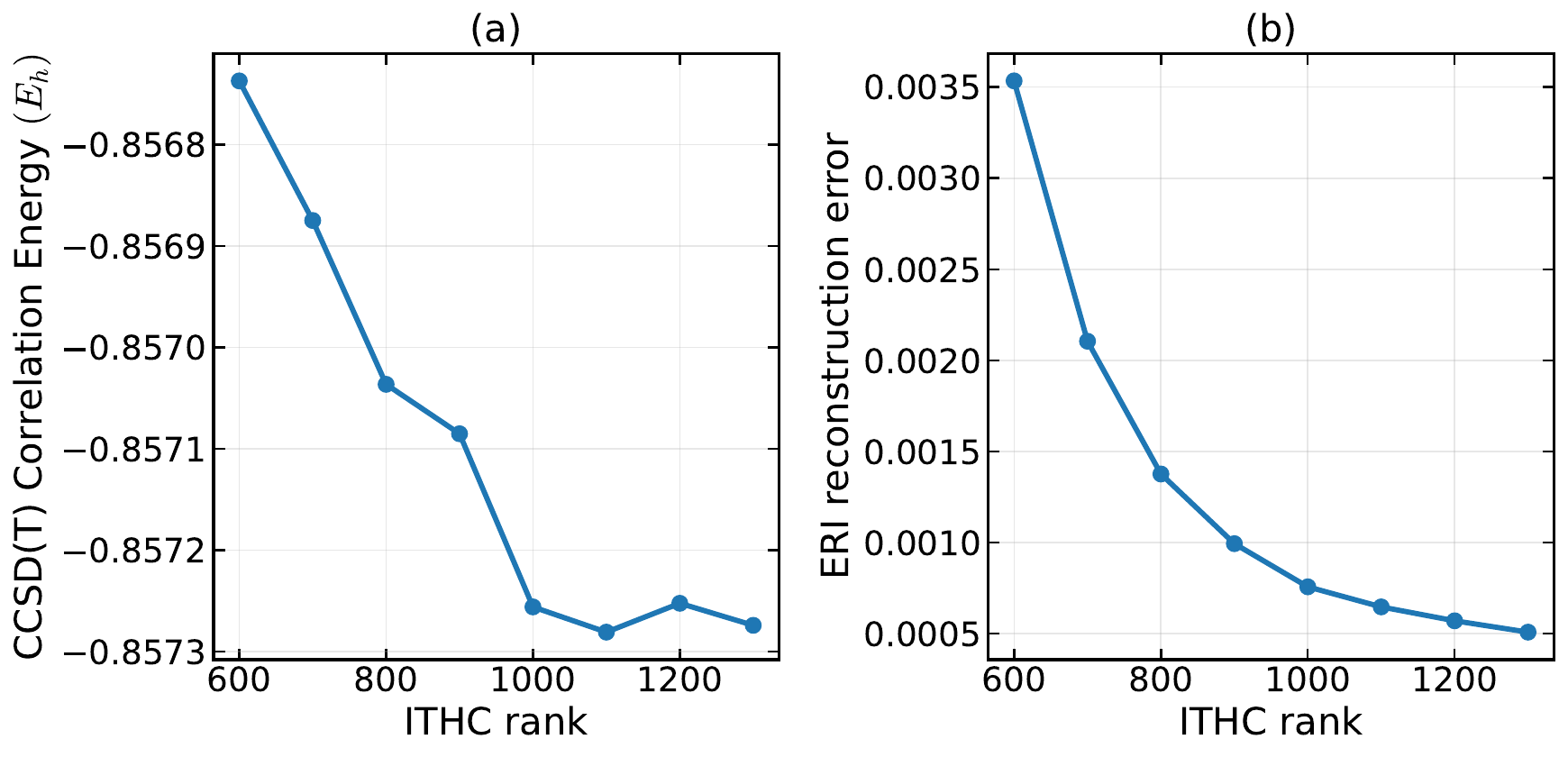}
    \caption{\textbf{(a)} CCSD(T) correlation energy of benzene in the cc-pVDZ basis with 30 electrons using ERIs reconstructed with different ITHC ranks and \textbf{(b)} ERI reconstruction error for different ITHC decomposition ranks. As the ITHC rank increases, the ERI reconstruction decreases, while the magnitude of the CCSD(T) correlation energy increases and approaches a converged value.}
    \label{fig:ITHC_CCSD}
\end{figure}

\changed{In order to quantify the effect the ITHC rank has on the AFQMC simulation, we investigated the accuracy of our ITHC-AFQMC method with respect to different ITHC ranks. For this, we ran ITHC-AFQMC simulations with varying ranks in the ITHC decomposition for the benzene molecule. 

First, to demonstrate the reliability of the ITHC decomposition itself, we computed the ERI reconstruction error for different ITCH ranks. We computed ITHC decompositions with ranks ranging from 600 to 1300. The results are shown in Fig.~\ref{fig:ITHC_CCSD} (a). The ERI reconstruction error decreases systematically as the ITHC rank increases, showing the systematic improvability of the ITHC method. We also performed deterministic CCSD(T) calculations on benzene, where the CCSD(T) correlation energy is shown in Fig.~\ref{fig:ITHC_CCSD} (b). In Fig.~\ref{fig:ITHC_CCSD} (b) the CCSD(T) correlation energy decreases with increasing ITHC rank, approaching the benchmark value given in Table~\ref{tab:results benzene}. This behavior demonstrates that the quality of the reconstructed ERIs systematically improves with increasing ITHC rank, thereby improving the quality of the CCSD(T) results.

To determine the effects of ITHC rank on the AFQMC results, we perform 10 independent AFQMC simulations for each rank, using different random seeds. The time step is set to $0.001$, and the energy estimate is obtained by averaging over the interval $\tau = 5-10$ a.u. The results are presented in Fig.~\ref{fig:ITHC_AFQMC}. Points of the same color correspond to individual simulations of the same rank, while the larger blue symbols denote the average energy over all seeds. The error bars represent the standard error of the mean, and the shaded regions indicate the 95\% confidence intervals.

For the AFQMC correlation energy (the blue symbols) in Fig.~\ref{fig:ITHC_AFQMC}, we observe greater variation for lower ITHC ranks (600-800). As the ITHC rank increases (900-1300), the energy variations become smaller, and a clear convergence trend emerges. These results indicate a systematic improvement of AFQMC energies with increasing ITHC rank. However, we note that the variation between independent AFQMC simulations with different random seeds is considerably large. This suggests a large stochastic uncertainty inherent to AFQMC. }

\changed{Overall, these results indicate that the results of AFQMC simulations can be systematically improved by increasing the ITHC rank. This is also corroborated by the deterministic CCSD(T) calculations, which reveal the expected convergence with increasing ITHC rank, thereby demonstrating the controllability of the ITHC approximation.}

\newpage 
\section{Conclusion and Outlook}

In this work, we presented a novel AFQMC method that reduces both theoretical complexity and practical runtime by diagonalizing the ERI in an extended space with additional modes. The resulting Hubbard-like two-body interaction term is then HS-transformed and used to perform imaginary-time propagation in an AFQMC simulation.

We integrated this new method into the existing ipie package and tested its efficacy for the hydrogen chain and the benzene molecule. For the $\ce{H10}$-chain, \changed{ITHC-AFQMC} recovered the ground-state energy within chemical accuracy relative to the FCI result, \changed{while exhibiting improved accuracy and numerical stability compared to the standard method}. For benzene, we recovered $99.8\%$ of the total correlation energy. Timing benchmarks further confirm that our approach yields superior GPU performance in both propagation and local energy estimation. While runtimes are comparable for small systems, the favorable memory usage of our method, $\mathcal{O}(N_{\mathrm{aux}}^2)$, provides a clear advantage as system size increases. \changed{Finally, we demonstrated the stability and controllability of the ITHC-AFQMC approach by showing the convergence of the benzene correlation energy with respect to the ITHC rank.}

\changed{We also performed intermediate method simulations, where the ITHC factors are used as Cholesky vectors within the standard AFQMC workflow. The numerical results for hydrogen chains showed that this intermediate method yields accuracy, numerical stability and runtimes comparable to the standard method. This demonstrates that the advantage of our method arises from the enlarged-basis framework rather than the ITHC decomposition alone.}

Recently, Xu \textit{et al.}~\cite{Xu.2024} applied a combination of DMRG and AFQMC to describe superconductivity in the 2D-doped Fermi-Hubbard model, demonstrating AFQMC's ability to treat strongly correlated systems. However, to fully capture the electronic correlation in strongly correlated systems, the transition to multi-determinant trial wavefunctions is essential. Since single Slater-determinants are exact solutions to non-interacting one-body problems, they inherently struggle to describe the multi-configurational nature and static correlation in complex systems~\cite{Lee.2020}. While the authors of the ipie package have demonstrated that using simple multi-determinant trial states can significantly improve accuracy, yielding results comparable to MBE-FCI, this approach introduces new challenges~\cite{Lee.2020,Malone.2023}. Specifically, the requirement of a CASSCF calculation to generate these determinants imposes a severe computational bottleneck, potentially limiting the scalability of AFQMC for the large-scale simulations that QMC methods are designed to enable~\cite{Lee.2020}.

Apart from multi-determinant trial states, there are several possible directions for further improving our method. At present, the error in each propagation step scales as \(\mathcal{O}(\Delta \tau^2)\), with the dominant contribution arising from the Zeno error. In previous work on quantum algorithms~\cite{Luo.2025}, the authors proposed a simple technique that reduces this error to \(\mathcal{O}(\Delta \tau^3)\). However, it is unclear how to adapt this technique to AFQMC while preserving the same favorable computational scaling.
\changed{Moreover, in the current ITHC workflow, although the ISDF procedure ensures a controllable initial guess, the final step still involves heuristic nonlinear optimization, which provides limited control over convergence and factorization quality. By contrast, standard Cholesky decomposition and ISDF-based THC offer more systematic and controllable procedures. Developing more robust and reliable algorithms for constructing the ITHC factors is, therefore, an important direction for future work. In Appendix~\ref{app:isdf-ithc}, we briefly discuss the possibility of constructing ITHC factors directly using ISDF.}

\subsection*{Data availability}

The ITHC data, the integrals for the systems studied in this work, and the data that support the findings of this paper are openly available in Ref.~\cite{zenodo_19368252}. The code is openly available on the ipie GitHub repository by by Joonho Lee \textit{et al.}~\cite{Malone.2023, jiang2024improved}.

\begin{acknowledgement}
We thank Ke Liao and Matthias Knechtges for inspiring discussions.
The research is funded by THEQUCO as part of the Munich Quantum Valley, which is supported by the Bavarian State Government with funds from the Hightech Agenda Bayern Plus. 
\end{acknowledgement}
\newpage
\appendix


\section{Pre-processing performance}
\label{app:metric}
\subsection{Three-step method}

Here, we briefly review the method for calculating the ITHC factors \cite{Luo.2025,luo_isometric_thc_github}, and report the computation time and performance metrics of the ITHC factors used in our extended AFQMC method in the main text.

As detailed in \cite{Luo.2025}, the ITHC pre-calculation consists of the following three steps:
\begin{enumerate}
    \item Computing THC.
    We first construct a generic tensor hypercontraction (THC) representation,
    \begin{align}
    V_{pqrs}
    =
    \sum_{\alpha,\beta=1}^{N_{\text{aux}}}
    X_{p\alpha} X_{q\alpha}
    W'_{\alpha\beta}
    X_{r\beta} X_{s\beta},
    \end{align}
    using the interpolative separable density fitting (ISDF) approach. The cutoff threshold in this algorithm is set to $10^{-5}$.
    \\ Remark:
    We find that a subsequent normalization,
    $
    X'_{p\alpha}
    =
    \frac{X_{p\alpha}}{\sqrt{ \sum_p |X_{p\alpha}|^2}},
    $
    can improve numerical stability. To further truncate the THC rank, one may perform pivoted QR decomposition on the matrix
    \begin{align}
    C_{\alpha\gamma}
    =
    \sum_{pq}
    X'_{p\alpha} X'_{q\alpha} L_{pq\gamma}, \quad \text{where}\; \sum_\gamma L_{pq\gamma} L_{rs\gamma} = V_{pqrs}
    \end{align}
    and retain the $\alpha$ indices with higher importance. Here, $L_{pq\gamma}$ could be obtained via Cholesky decomposition or singular value decomposition.
    \item 
     Convert $X'_{p\alpha}$ into an isometry $u_{p\alpha}$ = $\mu_{\alpha}X_{p\alpha}$. The scaling factor $\mu_\alpha$ is determined by solving the isometric condition
     \begin{align}
         \sum_{\alpha} u^*_{p\alpha} u_{q\alpha} = \sum_{\alpha} X_{p\alpha} X_{q \alpha} |\mu_\alpha|^2 = \delta_{pq}
         \label{eq:isometry_condition}
     \end{align}
     under positivity constraints. Specifically, we minimize
    \begin{align}
        \sum_{p,q} \left(\sum_{\alpha} X_{p\alpha} X_{q \alpha} |\mu_\alpha|^2 - \delta_{pq}\right)^2
    \end{align}
    subject to the condition $|\mu_\alpha |^2\geq \delta = 0.3$, using the convex optimization library Cvxpy \cite{diamond2016cvxpy, agrawal2018rewriting}.
    \\ Remark:  The existence of an approximate solution for Eq. \eqref{eq:isometry_condition} is proven in \cite{Luo.2025}, and follows from the connection between THC and separable density fitting. 
    \item Fine-tune $u_{i\alpha}$. We minimize $\epsilon_{\mathrm{ithc}}$ 
    \begin{equation}
    \epsilon_{\mathrm{ithc}} = \frac{\sum_{p,q,r,s} \abs{ \sum_{\alpha,\beta} u_{p\alpha} u_{q\alpha} W_{\alpha\beta} u_{r\beta} u_{s\beta}  - V_{pqrs}}^2}{\sum_{p,q,r,s} \abs{V_{pqrs}}^2},
    \end{equation}
    where $W_{\alpha \beta}$ is determined through
    \begin{align}
        W_{\alpha \beta} = u_{pq,\alpha}^{-1} V_{pq,rs} u_{rs,\beta}^{-1}, \quad u_{pq,\alpha}\equiv u_{p\alpha}u_{q\alpha}.
    \end{align}
    Here $u_{pq,\alpha}^{-1}$ implies the pseudo-inverse. 
    The optimization is performed using the Adam optimizer implemented in the Optax library~\cite{optax2020github}. We perform $1000$ optimization steps using an exponentially decaying learning-rate schedule with an initial learning rate of $10^{-3}$, a transition period of $1000$ steps, and a decay rate of $0.1$. We find that these parameter choices generally yield reliable convergence.
\end{enumerate}

We report the computational time required for each step of the algorithm for the systems studied in the main text, as shown in Tables \ref{tab:factors_hchain} and \ref{tab:factors_benzene}. Steps 1 and 2 were performed using 10 cores of an Intel Xeon CPU 8360Y at 2.40 GHz. Step 3 was performed on a single NVIDIA Tesla V100 GPU. Notice that the ITHC precomputation typically requires approximately 100 seconds. This timescale is much shorter than the wall time required for the AFQMC simulations, which is typically tens of minutes for non-trivial systems like benzene.

We also listed the detailed information on the ITHC factors used in our extended AFQMC method and the Cholesky factors used in the standard AFQMC method. The Cholesky decomposition is performed with the integrated module in the ipie library~\cite{Malone.2023}. 
Several quantities are considered. \(N_{\mathrm{chol}}\) and \(N_{\mathrm{ithc}}\) denote the ranks of the Cholesky decomposition and the ITHC factorization, respectively. They correspond to the number of auxiliary fields \(N_{\mathrm{aux}}\) in AFQMC. \(\epsilon_{\mathrm{chol}}\) and \(\epsilon_{\mathrm{ithc}}\) denote the relative errors of the Cholesky- and ITHC-approximated electron repulsion integral (ERI) tensor, \(V'\), evaluated using the Frobenius norm:
\begin{equation}
    \epsilon = \frac{\sum_{p,q,r,s} \abs{V'_{pqrs} - V_{pqrs}}^2}{\sum_{p,q,r,s} \abs{V_{pqrs}}^2}.
\end{equation}
We also evaluate the maximum norm of \(W_{\alpha\beta}\) in the ITHC representation, as larger values generally indicate larger Zeno errors~\cite{Luo.2025}. Overall, the ITHC exhibits a larger approximation error than the Cholesky decomposition.

\begin{table}[H]
\renewcommand{\arraystretch}{1.25}
\setlength{\tabcolsep}{1mm}
\begin{tabular}{c|r|c|r|c|c|c|c|c}
& $N_{\text{chol}}$ & $\epsilon_{\text{chol}}$ & $N_{\text{ithc}}$ & $\epsilon_{\text{ithc}}$ & $\max(W_{\alpha\beta})$   & Step 1 (s) & Step 2 (s) & Step 3 (s) \\ \hline
$H_{10}$ & 27  & $1.83 \cdot 10^{-7}$  & 27   & $7.94 \cdot 10^{-5}$        & 15.4 & 0.24 & 0.03 & 75.75 \\
$H_{20}$ & 57  & $2.16 \cdot 10^{-7}$  & 57   & $8.82 \cdot 10^{-5}$        & 16.0 & 1.93 & 0.05 & 123.43 \\
$H_{30}$ & 87  & $2.25 \cdot 10^{-7}$  & 87   & $6.74 \cdot 10^{-5}$        & 12.5 & 13.47 & 0.17 & 208.84 \\
$H_{40}$ & 117 & $2.28 \cdot 10^{-7}$  & 117  & $6.92 \cdot 10^{-5}$        & 5.4 & 28.81 & 0.26 & 219.87 \\
$H_{50}$ & 147 & $2.30 \cdot 10^{-7}$  & 147  & $7.40 \cdot 10^{-5}$        & 5.2 & 50.44 & 0.52 & 231.99 \\
$H_{60}$ & 177 & $2.32 \cdot 10^{-7}$  & 177  & $7.95 \cdot 10^{-5}$        & 4.9 & 81.74 & 0.87 & 247.43 \\
$H_{70}$ & 207 & $2.32 \cdot 10^{-7}$  & 207  & $8.53 \cdot 10^{-5}$        & 4.6 & 126.10 & 1.49 & 277.90 \\
$H_{80}$ & 237 & $2.33 \cdot 10^{-7}$  & 237  & $8.97 \cdot 10^{-5}$        & 5.4 & 193.93 & 2.52 & 334.74 
\end{tabular}
\caption{The performance metric of the Cholesky decomposition and ITHC for hydrogen chains. }
\label{tab:factors_hchain}
\end{table}

\begin{table}[H]
\renewcommand{\arraystretch}{1.25}
\setlength{\tabcolsep}{1mm}
\begin{tabular}{r|c|r|c|c|c|c|c}
$N_{\text{chol}}$ & $\epsilon_{\text{chol}}$ & $N_{\text{ithc}}$ & $\epsilon_{\text{ithc}}$ & $\max(W_{\alpha\beta})$ & Step 1 (s) & Step 2 (s) & Step 3 (s)\\ \hline
693 & $1.16\cdot 10^{-5}$ & 600  & $3.53 \times 10^{-3}$ & $9.6$ & 44.92 & 23.89 & 445.02 \\
 & & 700  & $2.11 \times 10^{-3}$ & $9.5$ & & 32.41 & 476.39 \\
 & & 800  & $1.38 \times 10^{-3}$ & $10.4$ & & 39.38 & 510.09 \\
 & & 900  & $9.94 \times 10^{-4}$ & $11.4$ & & 43.92 & 558.95 \\
 & & 1000 & $7.57 \times 10^{-4}$ & $11.2$ & & 53.73 & 605.28 \\
 & & 1100 & $6.47 \times 10^{-4}$ & $10.6$ & & 59.47 & 659.92 \\
 & & 1200 & $5.71 \times 10^{-4}$ & $13.3$ & & 73.93 & 714.71 \\
 & & 1300 & $5.08 \times 10^{-4}$ & $18.1$ & & 96.47 &
 765.28 \\
\end{tabular}
\caption{The performance metric of the Cholesky decomposition and ITHC for benzene. Step 1 is performed only once, after which the resulting factors are truncated to the desired ranks. }
\label{tab:factors_benzene}
\end{table}

\subsection{ISDF for ITHC}
\label{app:isdf-ithc}

As discussed in the previous section, the current three-step procedure for obtaining ITHC factors relies on heuristic nonlinear optimization, which provides only limited control over convergence and factorization quality. This motivates the question of whether ITHC factors can instead be obtained directly using the more controllable ISDF framework.

Given a real-valued orthonormal spatial orbital basis $\{ \chi_p \}$, the entries of the electron repulsion integral tensor are defined as
\begin{equation}
\label{eq:ERI_def}
V_{pqrs} = \int_{\mathbb{R}^3} \int_{\mathbb{R}^3} \chi_p(\mathbf{r}) \chi_q(\mathbf{r}) \frac{1}{\abs{\mathbf{r} - \mathbf{r}'}} \chi_r(\mathbf{r}') \chi_s(\mathbf{r}') \, d^3 r' \, d^3 r.
\end{equation}
In ISDF \cite{dong_interpolative_2018, lu_compression_2015}, the orbital products are approximated as
\begin{align}
\label{eq:isdf_approx}
    \chi_p(\mathbf{r}) \chi_q(\mathbf{r}) \approx \sum_{\mu=1}^{N_{\text{ISDF}}} \zeta_{\mu}(\mathbf{r}) \chi_p(\mathbf{x}_{\mu}) \chi_q (\mathbf{x}_{\mu}),
\end{align}
where the $\{\mathbf{x}_{\mu}\}$ can be interpreted as support points of the interpolating functions $\zeta_{\mu}$. \cite{lu_compression_2015}

Inserting Eq.~\eqref{eq:isdf_approx} into Eq.~\eqref{eq:ERI_def} leads to
\begin{equation}
\label{eq:thc_isdf}
V_{pqrs} \approx \sum_{\mu, \nu=1}^{N_{\text{ISDF}}} \chi_p(\mathbf{x}_{\mu}) \chi_q (\mathbf{x}_{\mu}) \, W_{\mu \nu} \, \chi_r(\mathbf{x}_{\nu}) \chi_s(\mathbf{x}_{\nu})
\end{equation}
with
\begin{equation}
W_{\mu \nu} = \int_{\mathbb{R}^3} \int_{\mathbb{R}^3} \frac{\zeta_{\mu}(\mathbf{r}) \zeta_{\nu}(\mathbf{r}')}{\abs{\mathbf{r} - \mathbf{r}'}} \, d^3 r' \,  d^3 r.
\end{equation}
As noted before~\cite{lu_compression_2015}, Eq.~\eqref{eq:thc_isdf} is already in THC format.

In the following, we aim to elucidate under which circumstances Eq.~\eqref{eq:thc_isdf} is actually of ITHC form, i.e., when do the matrices with entries $\chi_p(\mathbf{x}_{\mu})$ form isometries?
It turns out that it is sufficient to impose the following normalization condition on the interpolating functions:
\begin{equation}
\label{eq:isdf_normalisation_condition}
\int_{\mathbb{R}^3} \zeta_{\mu}(\mathbf{r}) \, d^3r = 1
\end{equation}
for all $\mu$. Namely, combining this normalization with Eq.~\eqref{eq:isdf_approx} leads to
\begin{equation}
\begin{aligned}
\sum_{\mu=1}^{N_{\text{ISDF}}} \chi_p (\mathbf{x}_{\mu}) \chi_q(\mathbf{x}_{\mu})%
= \int_{\mathbb{R}^3} \sum_{\mu=1}^{N_{\text{ISDF}}} \zeta_{\mu}(\mathbf{r}) \chi_p (\mathbf{x}_{\mu}) \chi_q (\mathbf{x}_{\mu}) \, d^3r%
\approx \int_{\mathbb{R}^3} \chi_p (\mathbf{r}) \chi_q (\mathbf{r}) \, d^3 r = \delta_{pq}.
\end{aligned}
\end{equation}
The last equal sign follows from the orthonormalization of the spatial orbital basis $\{ \chi_p \}$.

Therefore, in principle, finding interpolation functions that satisfy condition \eqref{eq:isdf_normalisation_condition} will directly yield a valid ITHC representation. We remark that the condition can be relaxed to $\int \zeta_{\mu}(\mathbf{r}) \, d^3r = c$ for any $c > 0$ (independent of $\mu$), since constant factors can be reabsorbed into the kernel $W_{\mu\nu}$. A straightforward approach to satisfy this condition is to choose interpolation points on a regular grid and interpolation functions $\zeta_{\mu}$ which are translated copies of each other. One concrete example is provided by Gausslet basis functions on a Cartesian grid~\cite{gausslet}. We leave a further investigation along this direction for future work.

\section{Error bounds}
\label{app:errors}

In this Appendix, we derive the error bound for a single imaginary time step. The discrepancy between the exact and numerical evolution is decomposed into three terms.
\begin{equation}
\begin{split}
&\norm{\left(e^{-\Delta \tau \hat{H}} - e^{- \frac{\Delta \tau }{2} \hat{h}} \mathbf{P} e^{-\Delta \tau \hat{W}}\mathbf{P} e^{- \frac{\Delta \tau }{2} \hat{h}}  \right) \ket{\phi}} \\
&\leq \norm{(e^{-\Delta \tau \hat{H}} - e^{- \Delta \tau(\hat{h} + \mathbf{P} \hat{W} \mathbf{P})  }) \ket{\phi}} \\
&+ \norm{\left(e^{- \Delta \tau(\hat{h} + \mathbf{P} \hat{W} \mathbf{P})} - e^{- \frac{\Delta \tau }{2} \hat{h}} e^{-\Delta \tau \mathbf{P}\hat{W}\mathbf{P}}  e^{- \frac{\Delta \tau }{2} \hat{h}}\right) \ket{\phi}} \\
&+ \Big\|\Big(e^{- \frac{\Delta \tau }{2} \hat{h}}  e^{-\Delta \tau \mathbf{P}\hat{W}\mathbf{P}} e^{- \frac{\Delta \tau}{2} \hat{h}} \\& \qquad - e^{- \frac{\Delta \tau }{2} \hat{h}} \mathbf{P} e^{-\Delta \tau \hat{W}}\mathbf{P} e^{- \frac{\Delta \tau }{2} \hat{h}}\Big) \ket{\phi}\Big\| \\
&= \epsilon_{\mathrm{ithc}} + \epsilon_{\mathrm{trotter}} + \epsilon_{\mathrm{zeno}}.
\end{split}
\end{equation}
These three terms correspond to the ITHC approximation ($\epsilon_{\mathrm{ithc}}$), Trotterization ($\epsilon_{\mathrm{trotter}}$), and quantum Zeno dynamics ($\epsilon_{\mathrm{zeno}}$), respectively. 

The ITHC error can be reduced by introducing a higher ITHC rank, and its quality is studied in the appendix \ref{app:metric}. The Trotter error is well studied and has been shown to scale as $\mathcal{O}(\Delta \tau^3)$~\cite{Childs_trotter}. The Zeno error, given that the target state $\ket{\phi}$ is in the space of $\mathbf{P}$, can be expressed as
\begin{equation}
\begin{split}
&\epsilon_{\mathrm{zeno}} \\
&= \Big\| \Big(e^{- \frac{\Delta \tau }{2} \hat{h}}  e^{-\Delta \tau \mathbf{P}\hat{W}\mathbf{P}} \mathbf{P}  e^{- \frac{\Delta \tau }{2} \hat{h}} \\ &\qquad - e^{- \frac{\Delta \tau }{2} \hat{h}} \mathbf{P} e^{-\Delta \tau \hat{W}}\mathbf{P}  e^{- \frac{\Delta \tau }{2} \hat{h}}\Big) \ket{\phi} \Big\| \\
&\le \norm{e^{- \frac{\Delta \tau }{2} \hat{h}}} \cdot \norm{e^{-\Delta \tau \mathbf{P}\hat{W}\mathbf{P}} \mathbf{P} - \mathbf{P} e^{-\Delta \tau \hat{W}}\mathbf{P}} \cdot \norm{e^{- \frac{\Delta \tau }{2} \hat{h}}} \\
&= \norm{e^{-\Delta \tau \mathbf{P}\hat{W}\mathbf{P}} \mathbf{P} - \mathbf{P} e^{-\Delta \tau \hat{W}}\mathbf{P}},
\end{split}
\end{equation}
which scales as $\mathcal{O}(\Delta \tau ^2 )$ \cite{Burgarth2020quantumzenodynamics}.

\section{Intermediate method}
\label{app:Intermediate}

Here, we present the detailed derivation and implementation of the intermediate method presented in Sec.~\ref{sec: H10 chain} and~\ref{sec:Timing Benchmarks}. To precisely isolate the source of the practical gains, we have implemented and benchmarked an intermediate variant that uses the ITHC representation solely as an ERI compression tool. The compressed ERI is converted back into an effective Cholesky-like factorized form and then executed entirely within a standard, non-enlarged basis ipie AFQMC propagator. We take the form of $\hat{W}$ given in Eq.~(\ref{eq: cholesky decomp of W}) and apply an eigenvalue decomposition.
\begin{align*}
    W_{\alpha\beta} = \sum_{\gamma = 1}^{N_{\text{aux}}} w_{\gamma \alpha} w_{\gamma \beta},
\end{align*}
which allows us to rewrite the ERI into
\begin{align*}
        V_{pqrs} &= \sum_{\alpha,\beta=1}^{N_{\text{aux}}} \sum_{\gamma = 1}^{N_{\text{aux}}} u_{p\alpha} u_{q\alpha} w_{\gamma \alpha} w_{\gamma \beta} u_{r\beta} u_{s\beta}\\
        &=\sum_{\gamma = 1}^{N_{\text{aux}}} L_{\gamma}^{pq} L_{\gamma}^{rs},
\end{align*}
which is a Cholesky decomposition, and thus we can employ the Cholesky vectors $L_{\gamma}^{pq}$ and $L_{\gamma}^{rs}$ in the standard ipie propagator.

\newpage

\bibliography{org}

\begin{tocentry}
  \centering
 \includegraphics[width=8.5cm]{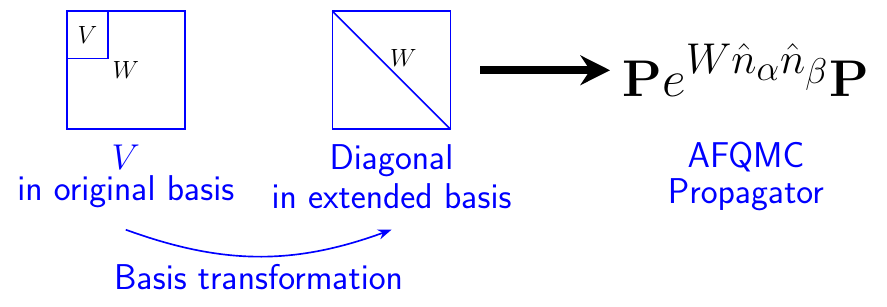}
 \label{TOC graphic}
\end{tocentry}

\end{document}